\documentclass[aps,prd,twocolumn,superscriptaddress]{revtex4-2}

\usepackage{graphicx}
\usepackage{dcolumn}
\usepackage{bm}
\usepackage{CJKutf8}

\usepackage[english,english]{babel}
\usepackage{amsmath}
\usepackage{amssymb,amsfonts,textcomp}
\usepackage{array}
\usepackage{supertabular}
\usepackage{multirow}
\usepackage{placeins}
\usepackage{hhline}
\usepackage[usenames]{color}
\usepackage{soul}
\usepackage{textcomp}
\usepackage{graphicx}
\usepackage{float}
\usepackage{dcolumn}
\usepackage{bm}
\usepackage{comment}
\usepackage[dvipsnames]{xcolor}
\usepackage[hidelinks]{hyperref}
\hypersetup{colorlinks=true, citecolor=RoyalBlue,linkcolor=RoyalBlue,filecolor=RoyalBlue, urlcolor=RoyalBlue}
\usepackage{float}
\usepackage{times}
\begin{document}
\begin{CJK*}{UTF8}{gbsn}

\preprint{PRD}

\title{Capturing primordial non-Gaussian signatures in the late Universe \\ by multi-scale extrema of the cosmic log-density field}

\author{Yun Wang (王云)}
\email{yunw@jlu.edu.cn}
\affiliation{College of Physics, Jilin University, Changchun 130012, China}

\author{Ping He (何平)}%
\email{hep@jlu.edu.cn}
\affiliation{College of Physics, Jilin University, Changchun 130012, China}%
\affiliation{Center for High Energy Physics, Peking University, Beijing 100871, China}%

\date{\today}

\begin{abstract}
We construct two new summary statistics, the scale-dependent peak height function (scale-PKHF) and the scale-dependent valley depth function (scale-VLYDF) of matter density, and forecast their constraining power on primordial non-Gaussianity and cosmological parameters based on \textsc{Quijote} and \textsc{Quijote-PNG} simulations at $z=0$. With the Fisher analysis, we demonstrate that these statistics outperform the power spectrum and bispectrum. Key findings include: (1) the constraint on the scalar spectral index $n_s$ obtained from the scale-VLYDF/scale-PKHF is 1.59/1.10 times tighter than that from the joint analysis of power spectrum and bispectrum; (2) the combination of the two statistics yields a slight improvement in constraining $\{f_\mathrm{NL}^\mathrm{local}, f_\mathrm{NL}^\mathrm{equil}\}$ over the power spectrum-bispectrum combination, and provides a 1.39-fold improvement in the constraint on $f_\mathrm{NL}^\mathrm{ortho}$; (3) after incorporating the power spectrum with our new statistics, parameter constraints surpass those from power spectrum-bispectrum combination by factors up to 2.93. This work offers an effective scheme for extracting primordial signals from the late Universe, paving the way for further breakthroughs in precision cosmology. 

\end{abstract}

\maketitle

\end{CJK*}

\section{Introduction}
\label{sec:intro}

The study of the early Universe is an essential topic in modern cosmology, with profound implications for the origin of the cosmos, the formation of cosmic structures, and fundamental physics. A critical aspect of this study is detecting and constraining the non-Gaussianity of primordial density fluctuations, i.e. \textit{primordial non-Gaussianity} (PNG),  which is a powerful probe to discriminate inflationary models, and to investigate the high energy physics of the early Universe \citep[see e.g.][for review]{Meerburg2019,Achucarro2022}.

The PNG is preserved throughout the evolution of cosmic matter distribution, leaving observable signatures in both the cosmic microwave background (CMB) and the large-scale structure (LSS) of the late Universe. To date, the most stringent constraints on the amplitudes of PNG, $f^\mathrm{X}_\mathrm{NL}$, come from measurements of the CMB anisotropies by the Planck satellite, which are $f^\mathrm{local}_\mathrm{NL}=-0.9\pm 5.1$, $f^\mathrm{equil}_\mathrm{NL}=-26\pm 47$, and $f^\mathrm{ortho}_\mathrm{NL}=-38\pm 24$ at $68\%$ C.L. \citep{Planck2020}, corresponding to local, equilateral, and orthogonal shapes of the primordial potential bispectrum, respectively \citep{Babich2004}. However, the two-dimensional (2D) nature and Silk damping hamper the further improvement of CMB's constraining ability \citep{Kalaja2021}. The ongoing and upcoming LSS surveys hold promise for offering enhanced sensitivity to PNG \citep{Alvarez2014,Karagiannis2018,Meerburg2019}, since they can map a huge three-dimensional volume of our Universe with high-scale resolution. Yet, this approach faces substantial complications due to the fact that feeble primordial information is obscured by the late-time non-Gaussianity induced by the non-linear gravity and other astrophysical processes.

Confronted with the challenge, the scientific community has persistently strived to develop sophisticated methodologies that go beyond the vanilla power spectrum and bispectrum of 3D density field, including but not limited to marked power spectrum \citep{Philcox2020,Massara2021,Massara2023}, power spectra in cosmic web environments \citep{Bonnaire2022,Bonnaire2023}, one-point probability distribution function \citep[PDF;][]{Uhlemann2018,Uhlemann2020,Friedrich2020}, neural network \citep{Giri2023,Floss2024}, persistent homology \citep{Biagetti2021,Biagetti2022,Yip2024}, and field-level inference \citep{Baumann2022,Andrews2023}. Which method optimizes the extraction of cosmological information from LSS remains an open question. 

Motivated by the above, we explore another potential avenue in this Letter, focusing on the following crucial features of the late-time matter distribution: First, the density field's PDF is nearly log-normal \citep{Hamilton1985,Coles1991}. Hence the logarithmic transform of the density field makes it more Gaussian-like and less non-linear \citep{Neyrinck2009,Wang2011,Rubira2021}. Second, the density field is manifested in a hierarchical web-like structure \citep{Sheth2004a,Sheth2004b,Shen2006}, which is most suitable to be analyzed with multi-scale tools, such as continuous wavelet transform \citep[CWT;][]{Daubechies1992,Romeo2008,Addison2017}. Third, the local extrema (halos/peaks and voids/valleys) of the density field have been shown to be particularly sensitive to the PNG \citep{Dalal2008,Chan2019}. Considering them all together, we first perform the wavelet transform of the log-density field, then identify local extrema on multiple scales and count them, thereby defining the \textit{scale-dependent peak height function} (scale-PKHF) and \textit{scale-dependent valley depth function} (scale-VLYDF). Here, we will demonstrate the outperformance of this pair of summary statistics in constraining PNG and standard cosmological parameters.

\begin{figure*}
\includegraphics[width=1\textwidth]{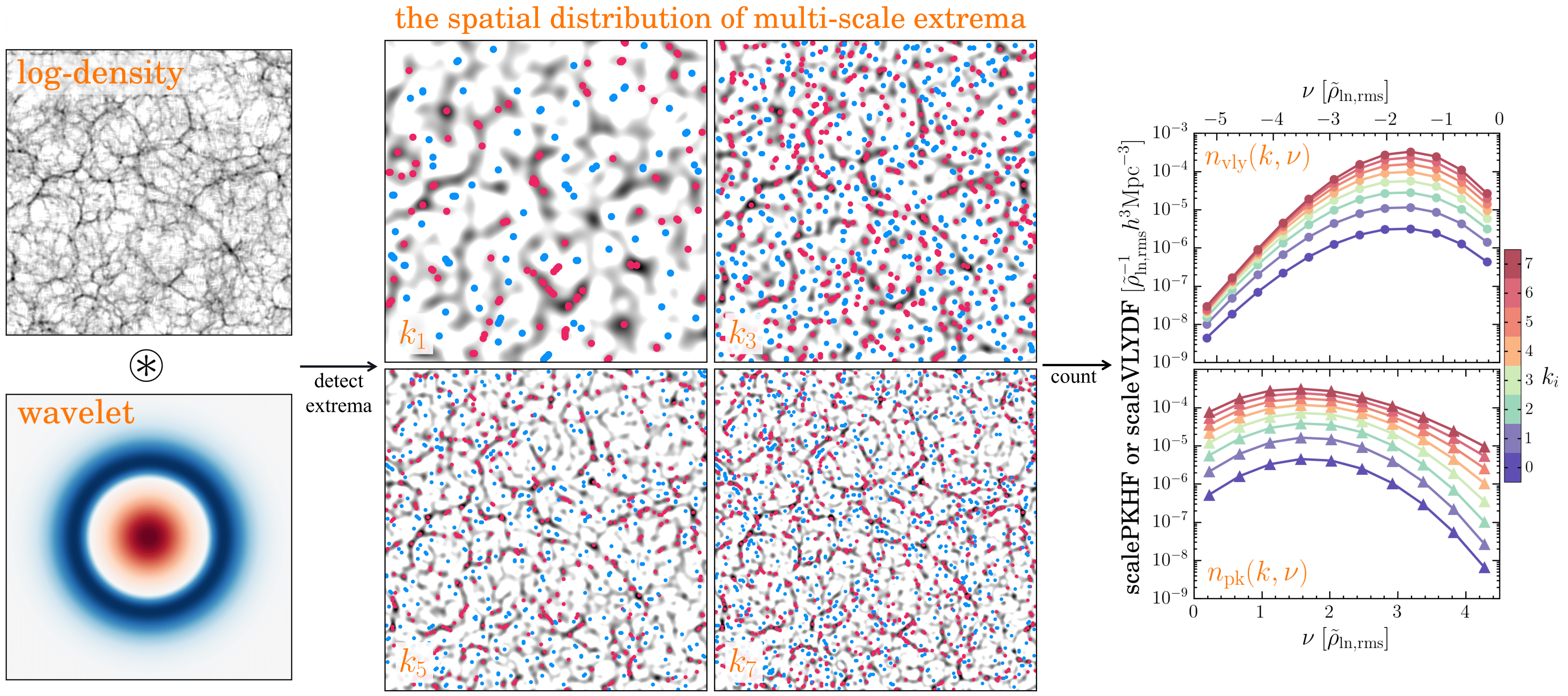}
\centering
\caption{\label{fig:scale_extrema} The illustration of measurements of the scale-PKHF and scale-VLYDF. The left panels represent the CWT operation, in which the 2D log-density slice of $500\times500\times20\ (h^{-1}\mathrm{Mpc})^3$ is drawn from a fiducial simulation of \textsc{Quijote} at $z=0$, and the 2D wavelet plot (with red for positive values and blue for negative values) is the cross-section of the isotropic GDW in X-Y plane. Here, the CWT operations are implemented on scales of $\{k_i|k_i\equiv w_i/c_w=(0.1+i\Delta_k) h\mathrm{Mpc}^{-1}\ \text{with}\ 0\leq i\leq 7\ \text{and}\ \Delta_k=2/35 \}$. Then we detect the local extrema of the CWT at each scale. For compactness, the middle panels show only the extrema of CWT fields on four selected scales, where gray regions indicate positive values of the CWT, blank regions indicate negative values, with red dots marking peaks and blue dots valleys. By counting the extrema, the measured scale-VLYDF $n_\mathrm{vly}(k,\nu)$ and scale-PKHF $n_\mathrm{pk}(k,\nu)$ are shown in the right panels. }
\end{figure*}

\begin{figure*}[t]
\includegraphics[width=0.95\textwidth]{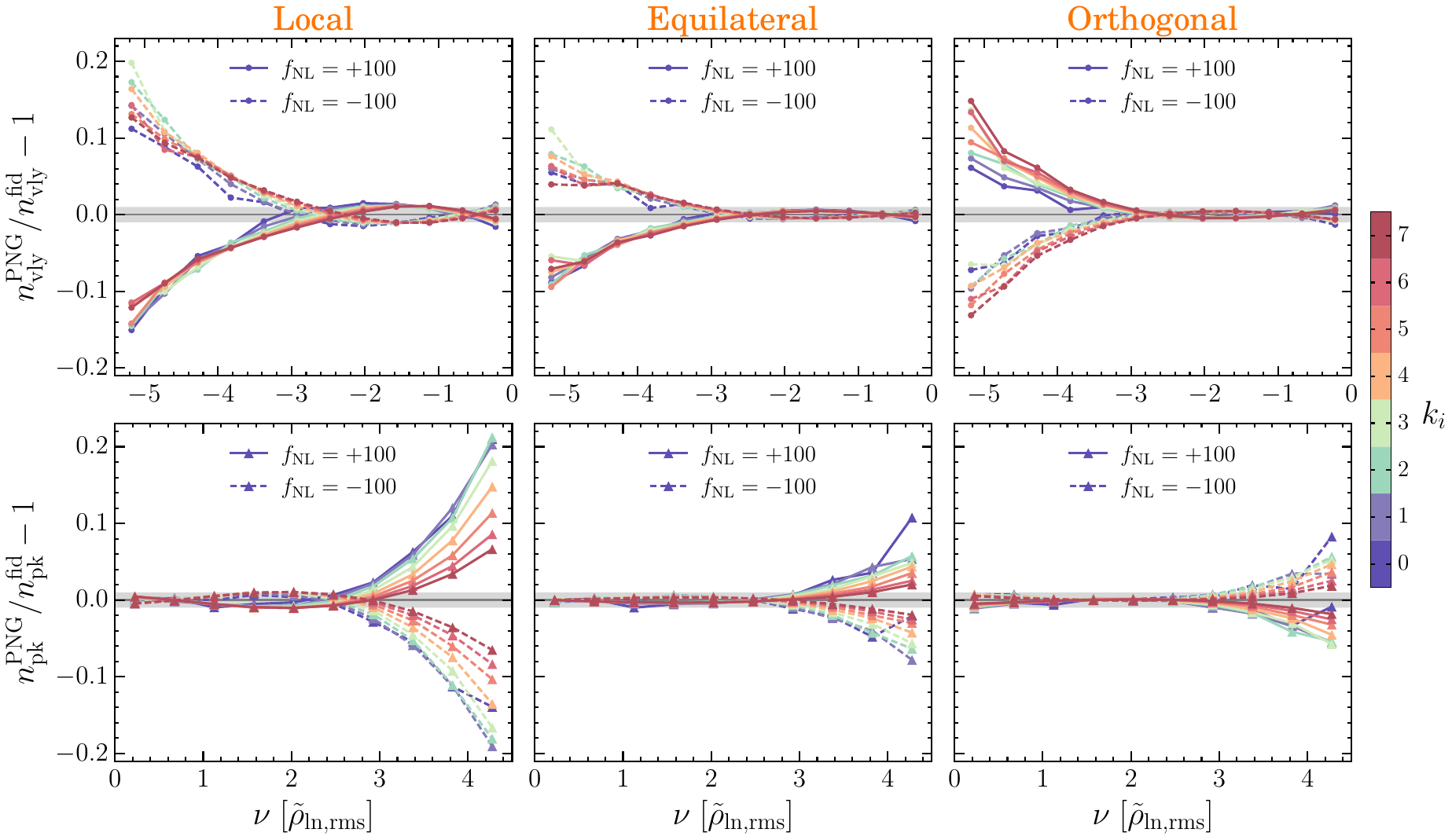}
\centering
\caption{\label{fig:png_effects} The impact of local (left), equilateral (middle), and orthogonal (right) PNG on the scale-VLYDF (top) and scale-PKHF (bottom). The superscript ``PNG" of the statistics indicates that they are averaged over $N_\mathrm{deriv}$ \textsc{Quijote-PNG} simulations, while the superscript ``fid" indicates that they are averaged over $N_\mathrm{fid}$ fiducial simulations. The solid lines with markers represent measurements for $f_\mathrm{NL}=+100$, and dashed lines with markers correspond measurements for $f_\mathrm{NL}=-100$. The grey bands denote the relative difference of 1 per cent.}
\end{figure*}

\section{Multi-scale extrema of the log-density}
\label{sec:scale_extrema}

With the aim of effectively mitigating the effects of late-time non-Gaussianity, we apply the logarithmic transform to the density field, which is given by
\begin{equation}
\label{eq:log_density}
    \rho_\mathrm{ln}(\mathbf{x})=\ln \left[1+\delta(\mathbf{x})\right],
\end{equation}
where the density field is constructed by assigning the particle positions to a regular grid with $N_\mathrm{g}=512^3$ cells using the piecewise cubic spline window function \citep{Sefusatti2016}. Then convolved with a wavelet $\Psi$, the CWT of the log-density field $\rho_\mathrm{ln}$ can be obtained as below
\begin{equation}
    \label{eq:cwt_log_density}
    \tilde\rho_\mathrm{ln}(w,\mathbf{x})=\int \rho_\mathrm{ln}(\mathbf{x}')\Psi(w, \mathbf{x}-\mathbf{x}')\mathrm{d}^3\mathbf{x}',
\end{equation}
in which $\Psi(w, \mathbf{x})=w^{3/2}\Psi(w\mathbf{x})$ is the rescaled wavelet of scale $w$. In the frame of CWT, there are numerous wavelet options \citep{Addison2017}. We use the isotropic Gaussian-derived wavelet \citep[GDW;][]{Wang2021,Wang2022a}, a Mexican-hat-shaped wavelet, as the mother wavelet,  given its suitability for detecting peaks and valleys across multiple scales \citep{Slezak1993}. Its explicit form is presented below
\begin{equation}
\label{eq:gdw}
\Psi(\mathbf{x}) = C_\mathrm{N}(6-|\mathbf{x}|^2)e^{-|\mathbf{x}|^2/4},
\end{equation}
where $C_\mathrm{N}=(15(2\pi)^{3/2})^{-1/2}$ is the normalization constant. For such isotropic wavelets, the log-density field can be reconstructed as follows \footnote{The derivation of Eq.\eqref{eq:inverse_cwt} can be referred to the one-dimensional case in Appendix A of \citep{Wang2023}.} 
\begin{equation}
    \label{eq:inverse_cwt}
    \rho_\mathrm{ln}(\mathbf{x}) = \langle \rho_\mathrm{ln} \rangle_\mathrm{V}+\frac{1}{\mathcal{K}_\Psi}\int_0^{+\infty}w^{\frac{1}{2}}\tilde\rho_\mathrm{ln}(w,\mathbf{x})\mathrm{d}w,
\end{equation}
where $\langle \rho_\mathrm{ln} \rangle_\mathrm{V}$ is the mean log-density over the whole space, and $\mathcal{K}_\Psi=\int_0^{+\infty}[\hat\Psi(k)/k]\mathrm{d}k$ with $\hat\Psi(k)$ being the Fourier transform of $\Psi(\mathbf{x})$. It can be seen from Eqs. \eqref{eq:cwt_log_density} and \eqref{eq:inverse_cwt} that the CWT provides a complete multi-scale picture of the matter distribution.

Next, we find the peaks/valleys of the CWT field $\tilde\rho_\mathrm{ln}(w,\mathbf{x})$ at a given scale $w$ by locating cells with values above/below their neighbors. By counting those extrema, we can define the scale-PKHF $n_\mathrm{pk}(w,\nu)$/scale-VLYDF $n_\mathrm{vly}(w,\nu)$ as the number density of CWT peaks/valleys with heights/depths falling in the bin $[\nu-\mathrm{d}\nu/2$, $\nu+\mathrm{d}\nu/2)$ per unit volume at scale $w$, which can be mathematically expressed as
\begin{equation}
\label{eq:scale_pkhf}
    n_\mathrm{pk}(w,\nu)=\frac{\mathrm{d}\mathcal{N}_\mathrm{pk}(w)}{\mathrm{d}\nu},
\end{equation}
and
\begin{equation}
\label{eq:scale_vlyhf}
    n_\mathrm{vly}(w,\nu)=\frac{\mathrm{d}\mathcal{N}_\mathrm{vly}(w)}{\mathrm{d}\nu},
\end{equation}
where $\mathcal{N}_\mathrm{pk}(w)$ and $\mathcal{N}_\mathrm{vly}(w)$ are the overall number densities of peaks and valleys at scale $w$, respectively. For comparing the scale-PKHF and scale-VLYDF with the power spectrum, we need to match the wavelet scale $w$ with the wavenumber $k$ by the correspondence $w=c_w k$ \citep[see Appendix A of][]{Wang2024b}, where $c_w=2/\sqrt{7}$ for the isotropic GDW. Then our measurements will be restricted to (i) 8 linearly spaced scales in the non-linear regime of $0.1\leq w/c_w\leq 0.5\ h\mathrm{Mpc}^{-1}$, (ii) 10 linear peak-height bins corresponding to $0<\nu\leq4.5\ \tilde\rho_{\ln,\mathrm{rms}}$, (iii) 12 linear valley-depth bins corresponding to $-5.4\ \tilde\rho_{\ln,\mathrm{rms}}\leq\nu<0$, where $\tilde\rho_{\ln,\mathrm{rms}}=\sqrt{ \langle|\tilde\rho_{\ln}(w,\mathbf{x})|^2\rangle_\mathrm{V} }$ denotes the mean square root at the scale $w$. With this configuration, the space is ensured to have at least one peak/valley on the largest scale and in the highest peak/deepest valley bin. Henceforth for convenience, we will use $k$ to replace $w$ without any ambiguity.

To gain physical intuition, we make a visualization of our basic idea and measurements in Fig. \ref{fig:scale_extrema}.

\section{Simulations}
\label{sec:simu}

The present work utilizes mock density fields at $z=0$ derived from the \textsc{Quijote} \citep{Villaescusa2020} and its extension \textsc{Quijote-PNG} \citep{Coulton2023a}, which are publicly available large suites of $N$-body simulations with a wide range of cosmological parameter space. Each simulation tracks the gravitational evolution of $512^3$ dark matter particles from $z=127$ to $z=0$ in a cubic box of side $L_\mathrm{box}=1 h^{-1}\mathrm{Gpc}$ using the TreePM code \textsc{Gadget-III} \citep{Springel2005}. The initial conditions of \textsc{Quijote} are Gaussian and generated by the 2LPTIC code \citep{Crocce2006}, while those of \textsc{Quijote-PNG} are non-Gaussian and generated by the 2LPTPNG code \citep{Scoccimarro2012,Coulton2023a}. 
The simulations are organized into different sets depending on their cosmological parameters. Among them, the fiducial set contains $N_\mathrm{fid}=15,000$ random realizations with parameters of $\{f^\mathrm{local}_\mathrm{NL}=0$, $f^\mathrm{equil}_\mathrm{NL}=0$, $f^\mathrm{ortho}_\mathrm{NL}=0$, $\Omega_m=0.3175$, $\Omega_b=0.049$, $\sigma_8=0.834$, $n_s=0.9624$, $h=0.6711\}$, which can be used to compute the covariance matrix of the statistic. Corresponding to each parameter, there is a simulation set containing $N_\mathrm{deriv}=500$ pairs of realizations, in which this parameter is perturbed by a small step around its fiducial value leaving the others unchanged. Then in this way, one can compute the partial derivative of the statistic concerning the parameter. For the parameters we considered here, the step sizes are $\{\mathrm{d}f^\mathrm{local}_\mathrm{NL}=\pm100$, $\mathrm{d}f^\mathrm{equil}_\mathrm{NL}=\pm100$, $\mathrm{d}f^\mathrm{ortho}_\mathrm{NL}=\pm100$, $\mathrm{d}\Omega_m=\pm0.010$, $\mathrm{d}\Omega_b=\pm0.002$, $\mathrm{d}\sigma_8=\pm0.015$, $\mathrm{d}n_s=\pm0.020$, $\mathrm{d}h=\pm0.020\}$.

\section{Fisher information analysis}
\label{sec:fisher}

The Fisher information matrix is a commonly used tool to assess the cosmological constraining power of a summary statistic \citep{Fisher1922,Tegmark1997,Carron2013}, which is defined as
\begin{equation}
    \label{eq:fisher_info}
    \mathcal{F}_{ij} = \left(\frac{\partial \langle\boldsymbol{S}\rangle_\mathrm{deriv}}{\partial \theta_i}\right)\mathcal{C}^{-1}\left(\frac{\partial\langle\boldsymbol{S}\rangle_\mathrm{deriv}}{\partial \theta_j}\right)^\mathrm{T},
\end{equation}
where $\langle\cdot\rangle_\mathrm{deriv}$ denotes the ensemble average over $N_\mathrm{deriv}$ paired simulations for each parameter. The statistic vector $\boldsymbol{S}$ is composed of the scale-PKHF, scale-VLYHF, and power spectrum, with elements ordered first by the scale $k$ and then by $\nu$ at each scale. $\theta_i$ is the $i$-th parameter of $\boldsymbol{\theta}=\{f^\mathrm{local}_\mathrm{NL}$, $f^\mathrm{equil}_\mathrm{NL}$, $f^\mathrm{ortho}_\mathrm{NL}$, $h$, $n_s$, $\Omega_m$, $\Omega_b$, $\sigma_8\}$, and $\mathcal{C}$ is the covariance matrix of statistic defined as
\begin{equation}
    \label{eq:cov}
    \mathcal{C}= \frac{1}{N_\mathrm{fid}-1}\sum_{n=1}^{N_\mathrm{fid}} (\boldsymbol{S}_n-\langle\boldsymbol{S}\rangle_\mathrm{fid})^\mathrm{T}(\boldsymbol{S}_n- \langle\boldsymbol{S}\rangle_\mathrm{fid}),
\end{equation}
in which the statistic $\boldsymbol{S}_n$ is measured from the $n$-th simulation, and $\langle\cdot\rangle_\mathrm{fid}$ denotes the ensemble average over $N_\mathrm{fid}$ fiducial simulations. To get an unbiased estimate, We multiply the inverse of the covariance matrix by the Hartlap factor of $(N_\mathrm{fid}-N_S-2)/(N_\mathrm{fid}-1)$ \citep{Hartlap2007}, where $N_S$ is the size of $\boldsymbol{S}$. 

The inverse of the Fisher matrix $\mathcal{F}^{-1}$ provides lower bounds on the parameter error covariance, with its diagonal elements estimating the  1-$\sigma$ marginalized error on parameters
\begin{equation}
    \label{eq:1sigma_error}
    \sigma^2(\theta_i)\geq (\mathcal{F}^{-1})_{ii}.
\end{equation}

\begin{figure}[t]
\centering
\includegraphics[width=0.95\columnwidth]{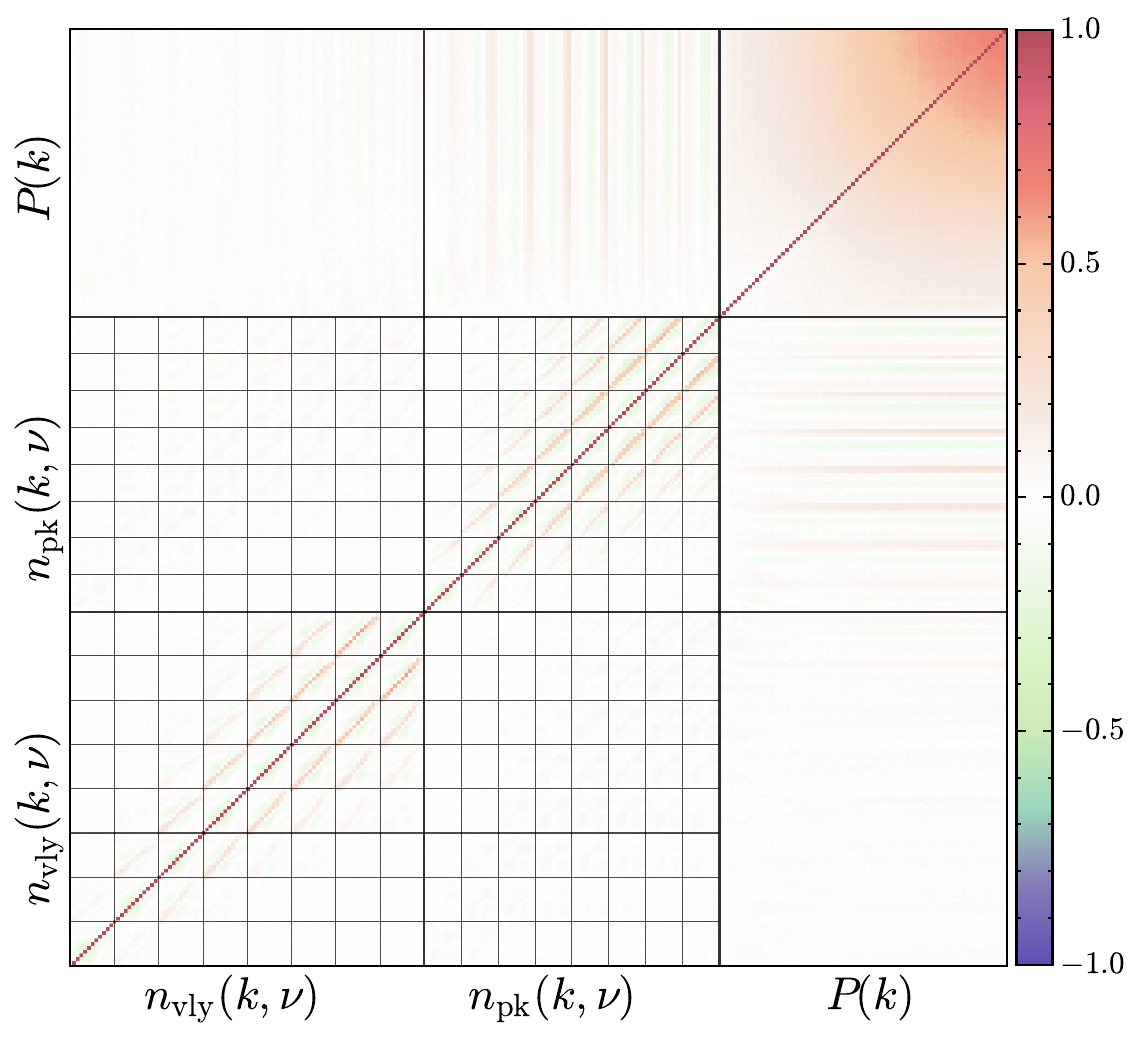}
\caption{\label{fig:correlation} The correlation matrix of the scale-VLYDF, scale-PKHF, and power spectrum. Note that for the two formers, gray lines partition the matrix into cells organized by scale. We compute the power spectrum by using the public Pylians3$^a$ library with wavenumber bins of width $k_F=2\pi/L_\mathrm{box}$, from $k_F$ to $0.5\ h\mathrm{Mpc}^{-1}$.\\
$^a$ \url{https://pylians3.readthedocs.io/en/master/}}
\end{figure}

\section{Results}
\label{sec:results}

We first investigate the impact of PNG on the scale-PKHF and scale-VLYDF by comparing the average of measurements over \textsc{Quijote-PNG} simulations and that over fiducial \textsc{Quijote} simulations, the results of which are presented in Fig. \ref{fig:png_effects}. We observe that all shapes of PNG can introduce sizable effects on both statistics, with magnitudes of $\gtrsim 1$ per cent for $|\nu|\gtrsim 3\tilde\rho_\mathrm{ln,rms}$. Positive and negative PNG parameters produce opposite impacts on the statistics. Specifically, local PNG of $f_\mathrm{NL}^\mathrm{local}=+100$ leads to a decrease in the number of deep valleys and an increase in the number of high peaks with magnitudes of $\sim10-20$ per cent for the largest extrema, whereas local PNG of $f_\mathrm{NL}^\mathrm{local}=-100$ does the reverse exactly. The impact of equilateral PNG is similar to it, but with a weaker magnitude. However, the orthogonal PNG has a different effect, e.g. the value of $f_\mathrm{NL}^\mathrm{ortho}=+100$ would increase the amount of deep valleys and decrease the amount of high-peaks. We also see that the PNG effects on scale-PKHF and scale-VLYDF vary with scale. The detailed study and theoretical modeling of these effects lie beyond the scope of this study and are left for future research. 

\begin{figure}[ht]
\centering
\includegraphics[width=0.86\columnwidth]{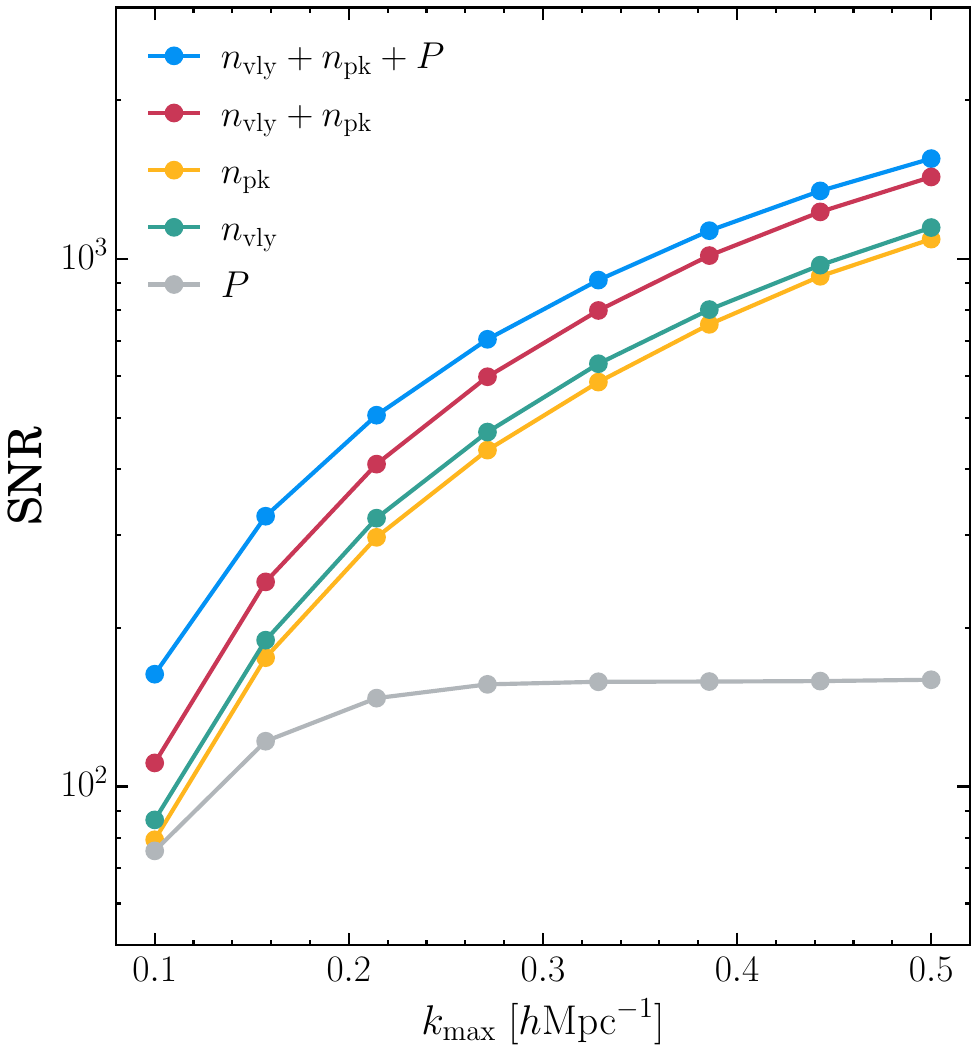}
\caption{\label{fig:snr} The cumulative SNR of the scale-VLYDF, scale-PKHF, power spectrum, and their combinations, as labeled. The eight maximum wavenumbers are actually the scales at which we perform the CWT.}
\end{figure}

A theoretical understanding of the covariance matrix of summary statistics is particularly important for parameter forecasting from surveys. For this, we show in Fig. \ref{fig:correlation} the normalized covariance (i.e. correlation) matrix $r_{ij}\equiv \mathcal{C}_{ij}/\sqrt{\mathcal{C}_{ii}\mathcal{C}_{jj}}$ of the scale-VLYDF, scale-PKHF, and power spectrum, computed numerically using Eq. \eqref{eq:cov}. As expected, the covariances of the scale-PKHF and scale-VLYDF are more diagonalized than those of the power spectrum, enabling more cosmological information to be retrieved. It can also be seen that there is a minimal correlation between those statistics, indicating that the information they provide is complementary.

\begin{figure*}
\includegraphics[width=0.9\textwidth]{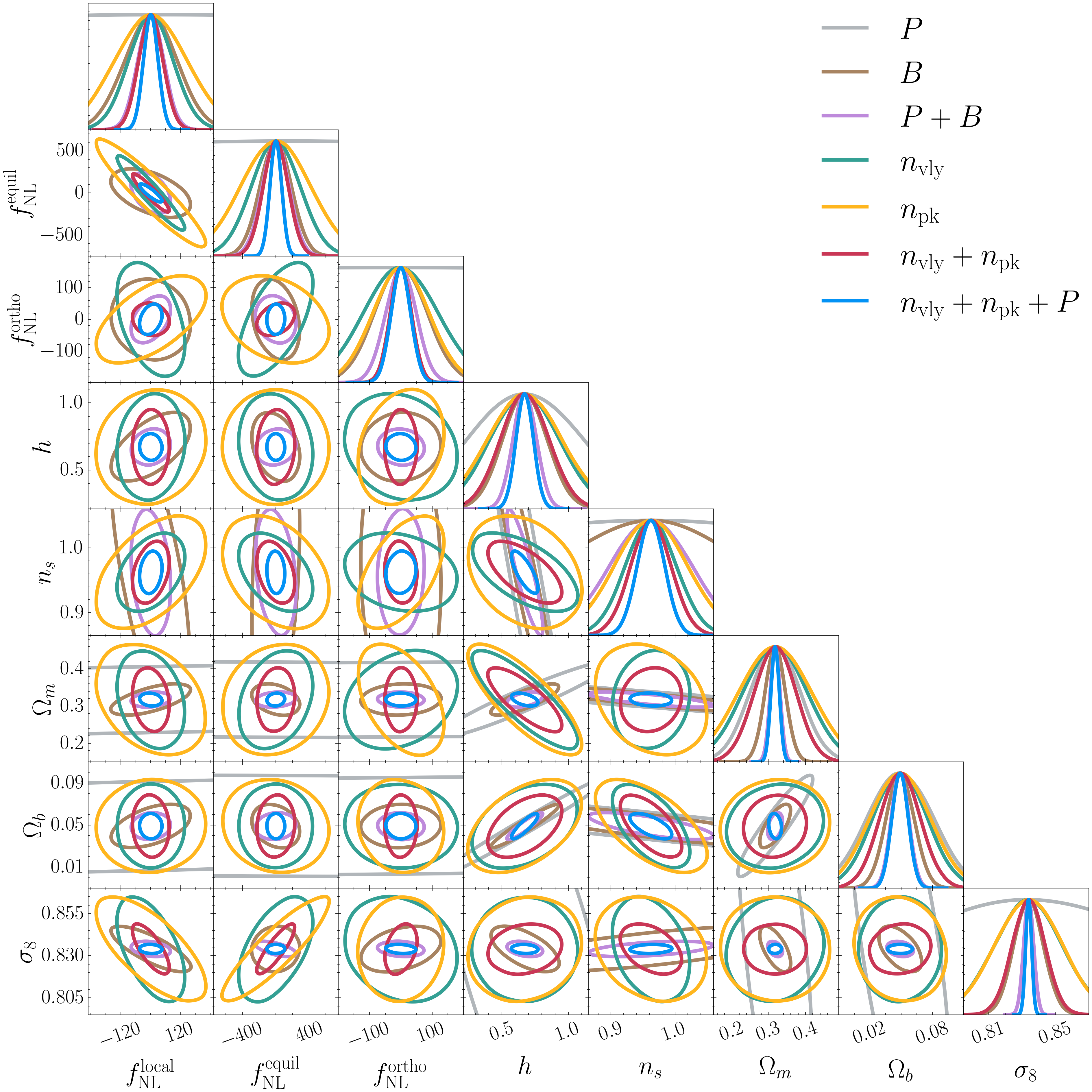}
\centering
\caption{\label{fig:likelihood} The marginalized 1-$\sigma$ confidence contours for PNG and standard cosmological parameters obtained from the scale-VLYDF, scale-PKHF, power spectrum, bispectrum, and some of their combinations, as labeled. In some subplots, the confidence ellipses from the power spectrum are too large to fit within the plotted range. The results for the bispectrum and its combination with power spectrum are sourced from \cite{Floss2024} and consistent with those presented in \cite{Coulton2023a}, which are the measurements with the maximum wavenumber being cut off at $0.5\ h\mathrm{Mpc}^{-1}$ and without applying the reconstruction algorithm.}
\end{figure*}

\begin{table*}[t]
\caption{\label{tab:improve_fac}%
The improvement factors of various statistics over the ordinary power spectrum for PNG amplitudes and cosmological parameters.}
\begin{ruledtabular}
\begin{tabular}{lcccccc}
\textrm{Paras}&
$\sigma_P/\sigma_B$ &
$\sigma_P/\sigma_{P+B}$ &
$\sigma_P/\sigma_{n_\mathrm{vly}}$ &
$\sigma_P/\sigma_{n_\mathrm{pk}}$ & 
$\sigma_P/\sigma_{n_\mathrm{vly}+n_\mathrm{pk}}$ &
$\sigma_P/\sigma_{n_\mathrm{vly}+n_\mathrm{pk}+P}$\\
\colrule
$f_\text{NL}^\text{local}$  & 28.59 & 57.61 & 32.73 & 20.22 & 60.24 & 99.14 \\
$f_\text{NL}^\text{equil}$ & 45.10 & 53.28 & 28.05 & 19.50 & 54.82 & 115.97 \\
$f_\text{NL}^\text{ortho}$ & 43.52 & 74.96 & 29.82 & 39.32 & 104.36 & 112.42 \\
$h$        & 2.59 & 4.94 & 1.66 & 1.53 & 2.36 & 6.59 \\
$n_s$      & 3.17 & 7.80 & 12.37 & 8.61 & 15.61 & 22.82 \\
$\Omega_m$ & 2.47 & 5.11 & 0.78 & 0.69 & 1.20 & 5.92  \\
$\Omega_b$ & 2.37 & 3.83 & 1.20 & 1.09 & 1.62 & 4.01  \\
$\sigma_8$ & 10.06 & 29.88 & 4.24 & 4.23 & 8.81 & 48.46
\end{tabular}
\end{ruledtabular}
\end{table*}

Given the covariance matrix, we can also determine the cumulative signal-to-noise ratio (SNR), which is another useful proxy for the information content of the summary statistic, defined as
\begin{equation}
    \label{eq:snr}
    \mathrm{SNR} = \sqrt{ \langle\boldsymbol{S}\rangle_\mathrm{fid}\mathcal{C}^{-1}\langle\boldsymbol{S}\rangle_\mathrm{fid}^\mathrm{T} }.
\end{equation}
Its estimations are displayed in Fig. \ref{fig:snr} for the studied statistics as a function of the maximum wavenumber $k_\mathrm{max}$. We see that the power spectrum SNR flattens out beyond $k\approx 0.3\ h\mathrm{Mpc}^{-1}$, which has also been reported in previous studies, and that the bispectrum follows the same feature with a lower SNR \citep{Chan2017,Bonnaire2022,Coulton2023a,Coulton2023b}. In contrast, both the scale-PKHF and scale-VLYDF do not experience such flattening and achieve a high SNR level. The combination of them both gives a much better SNR, up to 8.98 times higher than the power spectrum at $k_\mathrm{max}=0.5\ h\mathrm{Mpc}^{-1}$, and even 9.73 times when the power spectrum is included. We note that the combination of power spectra in cosmic web environments in \citep{Bonnaire2022} can achieve an 8 times higher SNR than the ordinary power spectrum.

In Fig. \ref{fig:likelihood}, we present the Fisher forecast for 1-$\sigma$ confidence contours of PNG and cosmological parameters at the maximum wavenumber $k_\mathrm{max}=0.5\ h\mathrm{Mpc}^{-1}$. Considering that the power spectrum carries negligible information on PNG, we also include the constraints from the bispectrum and its combination with the power spectrum for comparison \citep{Coulton2023a,Floss2024}. For further clarity, we also list explicitly the improvement factors over the power spectrum in Tab. \ref{tab:improve_fac} for all statistics. We see that the use of either scale-VLYDF or scale-PKHF alone can boost constraints significantly on the three types of PNG and break key degeneracies between parameters, including those between different PNG types, those between PNG amplitudes and cosmological parameters, and those between different cosmological parameters, e.g. $n_s-\sigma_8$, $h-\sigma_8$, $\Omega_b-\sigma_8$, and $n_s-h$. It is noteworthy that the scalar spectral index $n_s$ is constrained much more tightly by the scale-VLYDF and scale-PKHF, outperforming the power spectrum by factors of 12.37 and 8.61, the bispectrum by factors of 3.90 and 2.72, and the power spectrum-bispectrum combination by 1.59 and 1.10, respectively. However, they provide weaker constraints on $\Omega_m$ than that from the power spectrum.

After jointly combining the scale-PKHF and scale-VLYDF, all the parameter constraints become more stringent than those from the power spectrum. For the primordial parameter subset $\{f_\mathrm{NL}^\mathrm{local}, f_\mathrm{NL}^\mathrm{equil},f_\mathrm{NL}^\mathrm{ortho}, n_s\}$ embedding the primordial information, constraints are improved approximately by factors of $\{2.11, 1.22, 2.40, 4.92 \}$ over the bispectrum, and by $\{1.05, 1.03, 1.39, 2.00 \}$ relative to the bispectrum and power spectrum combination. When the power spectrum is included in our combination, all constraints are further improved, reaching $\{1.72, 2.18, 1.50, 1.33, 2.93, 1.16, 1.05, 1.62\}$ times the joint constraints from the power spectrum-bispectrum combination for parameters $\{f^\mathrm{local}_\mathrm{NL}$, $f^\mathrm{equil}_\mathrm{NL}$, $f^\mathrm{ortho}_\mathrm{NL}$, $h$, $n_s$, $\Omega_m$, $\Omega_b$, $\sigma_8\}$. This demonstrates that tighter parameter constraints can be achieved by combining the power spectrum with the scale-VLYDF and scale-PKHF,  bypassing the reliance on higher-order spectra (or correlation functions). 

\section{Conclusions}
\label{sec:conc}

In this Letter, we propose a pair of new summary statistics, the scale-PKHF and scale-VLYDF of 3D density field, which are well-defined, easy to implement, and fully leverage the multi-scale nature, log-normal property, and local extrema distribution of the matter distribution at late-times. Based on massive data sets of \textsc{Quijote} and \textsc{Quijote-PNG} simulations, we apply the two statistics to forecast the primordial non-Gaussianity and cosmology with the Fisher matrix formalism. 

We first observe that all shapes of PNG have significant effects on the scale-PKHF and scale-VLYDF, with magnitudes exceeding $1$ per cent for large extrema. Different shapes and amplitudes of PNG can lead to different impacts on the statistics, demonstrating the ability of these summary statistics to differentiate between various PNG models. 

We find that the covariance matrix of the scale-PKHF and scale-VLYDF shows less scale coupling than that of the ordinary power spectrum. Combining the scale-PKHF and scale-VLYDF with the power spectrum can achieve a high SNR of $\sim$10 times the power spectrum at the maximum wavenumber of $0.5\ h\mathrm{Mpc}^{-1}$ without showing signs of flattening. These facts suggest that the two statistics can extract huge information content from the LSS. Further, we note that the sole use of scale-VLYDF or scale-PKHF can already put a tighter constraint on the scalar spectral index $n_s$ than the power spectrum-bispectrum combination. Jointly considering the scale-PKHF and scale-VLYDF leads to much stronger constraints on the amplitudes of PNG $\{f_\mathrm{NL}^\mathrm{local}, f_\mathrm{NL}^\mathrm{equil},f_\mathrm{NL}^\mathrm{ortho} \}$ than the bispectrum,  while offering a modest improvement over the power spectrum-bispectrum combination, which highlights that the scale-PKHF and scale-VLYDF are very sensitive to the faint primordial signals in the LSS. By incorporating the power spectrum, scale-PKHF, and scale-VLYDF, all parameter constraints are tightened compared to those from the power spectrum-bispectrum combination. The greatest improvement is observed for $n_s$, followed by the PNG parameters and $\sigma_8$, with the least improvement for the remaining parameters.

Overall, we conclude that our methodology shows great superiority in constraining PNG and cosmological parameters. 
Notably, next-generation surveys like DESI \citep{DESI2016} and Euclid \citep{Amendola2018}, which aim to map the Universe at unprecedented resolution, can benefit significantly from these new statistics. By leveraging their ability to extract complementary information from the LSS, these surveys can achieve tighter constraints on PNG and other cosmological parameters. Furthermore, the enhanced sensitivity to the scalar spectral index underscores the potential of our methods to refine our understanding of inflationary physics. Nonetheless, it is important to recognize that our analysis is confined to the dark matter density field, which cannot be observed directly and is instead traced by galaxies. Observational effects, such as galaxy bias, redshift-space distortions, survey geometry, and selection functions, can degrade the parameter constraints. To adapt our statistics for real data from LSS surveys and achieve optimal constraints, we plan to leverage the \textsc{SimBIG} \citep{Hahn2022,Hahn2024} framework to produce galaxy mock catalogs, incorporating observational effects to robustly infer the posterior distribution of parameters. 

We release our code for reproducing our results at \url{https://github.com/WangYun1995/CWTextrema}.

\section*{Acknowledgments}

We would like to acknowledge the anonymous referee for valuable comments and suggestions.
We are grateful to Thomas Fl{\"o}ss for providing parameter constraints estimated from the power spectrum, bispectrum, and their combination. We also thank Francisco Villaescusa-Navarro, William Coulton, and the whole \textsc{Quijote} team for making their simulations publicly available. This work was supported by the National Science Foundation of China (Nos. 12147217, 12347163), the China Postdoctoral Science Foundation (No. 2024M761110), and the Natural Science Foundation of Jilin Province, China (No. 20180101228JC).

\nocite{*}
\bibliography{paper}

\end{document}